\documentclass[%
preprint,%
 aip,  
 amssymb, amsmath,%
]{revtex4-1}

\usepackage{graphicx}%
\usepackage{epstopdf}
\usepackage{color}
\usepackage{bm}%
\usepackage{sidecap}
\sidecaptionvpos{figure}{t}

\begin{document}

\title{Asymmetric propagation using enhanced self-demodulation in a chirped phononic crystal}

\author{A. Cebrecos}
\email{alejandro.cebrecos@univ-lemans.fr}
\author{N. Jim\'enez}
\author{V. Romero-Garc\'ia}
\affiliation{LUNAM Universit\'e, Universit\'e du Maine, CNRS, LAUM UMR 6613, Av. O. Messiaen, 72085 Le Mans, France}

\author{R. Pic\'o}
\author{V. J. S\'anchez-Morcillo}
 \affiliation{Instituto de Investigaci\'on para la Gesti\'on Integrada de zonas Costeras, Universitat Polit\`ecnica de Val\`encia, Paranimf 1, 46730, Grao de Gandia, Val\`encia, Spain}

\author{L. M. Garcia-Raffi}
 \affiliation{Instituto Universitario de Matem\'atica Pura y Aplicada, 
 Universidad Polit\'ecnica de Valencia, Camino de Vera s/n, 46022, 
 Valencia, Spain}

\begin{abstract}
Asymmetric propagation of acoustic waves is theoretically reported in a chirped phononic crystal made of the combination of two different nonlinear solids. The dispersion of the system is spatially dependent and allows the rainbow trapping inside the structure. Nonlinearity is used to activate the self-demodulation effect, which is enhanced due to the particular dispersion characteristics of the system. The main feature of the device is that integrates the nonlinear and the filtering effects in a single propagating medium. The performed numerical study reveals an efficient generation of the demodulated wave, up to 15\% in terms of the pressure amplitude, as well as strong attenuation for undesired frequency components above the cut-off frequency. The obtained energy rectification ratio is in the order of $10^4$ for the whole range of amplitudes employed in this work, indicating the robustness of the asymmetry and non-reciprocity of the proposed device for a wide operational range.
\end{abstract}

\maketitle


\section{Introduction}
\label{sec:intro}

There is an increasing interest in non-reciprocal acoustic devices, those presenting an asymmetric propagation of waves due to the breaking of the time reversal invariance. In certain cases, such devices are able to transmit acoustic waves only in one direction. Inspired by their corresponding electrical counterparts, many interesting applications of this effect have been suggested, and the corresponding devices have been often called acoustic or phonon ‘diodes’, although there has been a little abuse with this denomination and must be used with care \cite{maznev2013reciprocity}. Phononics, the science of elastic/acoustic wave propagation in artificially periodic media, has been shown as an ideal platform for the design of acoustical nonreciprocal devices. Moreover, acoustic diodes have been proposed as one of the most promising applications of phononics \cite{maldovan2013sound}.

Once nonlinearity is considered in the wave propagation problem, it automatically introduces a plethora of rich wave phenomena in the system\cite{Hamilton98Ch}. It has been claimed that nonlinearity is needed to break reciprocity \cite{maznev2013reciprocity}. Nonlinear materials have the ability to change the spectrum of waves propagating in a medium, generating higher harmonics of a pure-tone input signal, or combination frequencies in the case of multi-frequency input. Original proposals of acoustic diodes exploiting nonlinear effects used highly nonlinear media, as artificial structures made of microbubbles \cite{liang2010acoustic} or granular chains\cite{boechler2011bifurcation}. In both cases, a nonlinear medium is located next to a periodic structure. When the wave passes first through the nonlinear medium, higher harmonics are generated, which can be blocked by the crystal if the bandgap is properly tuned. However, if we interchange input and output, the harmonics will be transmitted since the wave radiates the crystal in a propagating band and harmonics are generated later.
 	
More recently, an acoustic diode has been proposed based on the self-demodulation effect \cite{devaux2015asymmetric}. It manifests due to the nonlinear generation of a low-frequency signal by a pulsed and high-frequency sound wave (or alternatively due to the nonlinear mixing of two high-frequency carrier waves) which produces a low-frequency wave corresponding to the difference frequency. Therefore, the proposal is based on the creation of lower, instead of higher frequencies as in Ref. \onlinecite{liang2010acoustic}, although the diode itself is structurally similar.
 	
\begin{figure}[htbp]
	\includegraphics[width=14cm]{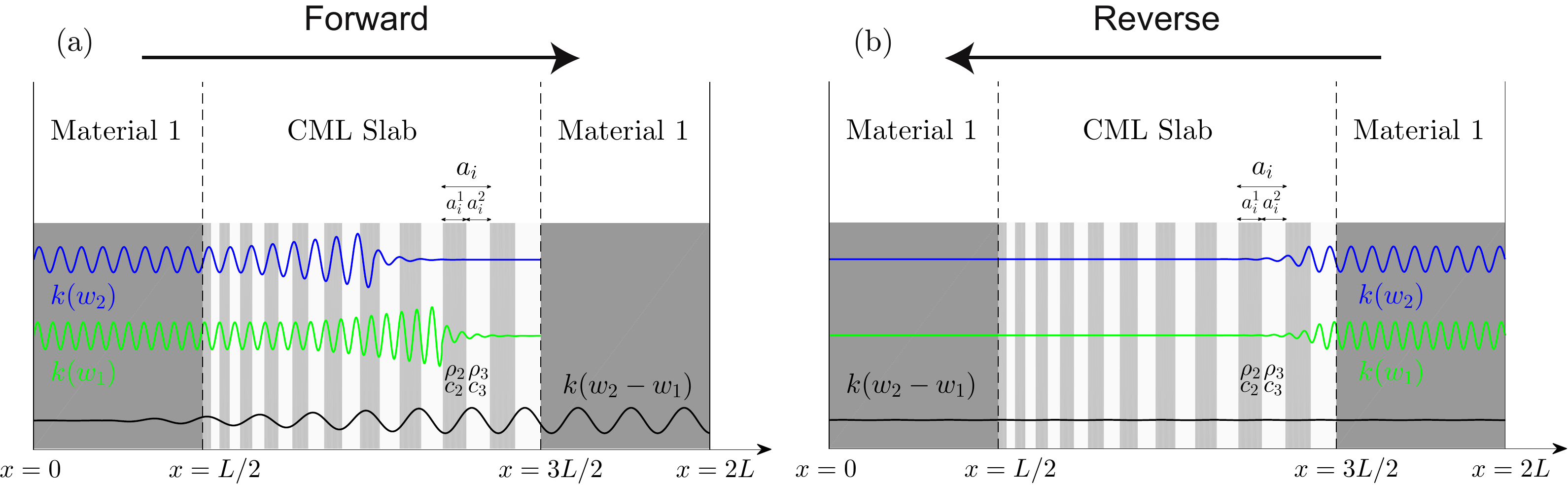}
	\caption{Scheme of the proposed system for asymmetric wave 
		propagation composed of a CPC structure placed in the middle of two homogeneous slabs of material 1. (a) Forward configuration. (b) Reverse configuration.} 
	\label{fig:fig1}
\end{figure}

  An interesting variation of conventional periodic crystals, which have been shown very useful for wave manipulation purposes, are the chirped (or graded) crystals, in which lattice constant is no more constant, but changes gradually along the propagation direction. Therefore, the dispersion properties change spatially along the structure and in some cases, they can be tailored to design a rainbow mirror\cite{Romero13, Cebrecos14}. In this case once the bandgap is reached, the wave is no more allowed to propagate and therefore is backward reflected with amplitude enhancement at different planes due to a progressive slowing down of the wave as it propagates along the structure. Many interesting wave-managing effects have been reported, as the mirage and superbending effect in bulk crystals \cite{Centeno06} and waveguides \cite{Wu11}, or the sub-wavelength rainbow trapping \cite{Shen11,Smolyaninova10,Zhu13}.  
Here we consider for the first time the propagation of sound or elastic waves in a chirped phononic crystal (a multilayered structure in the 1D case discussed here) in the  nonlinear regime, i.e., we assume that the material forming the crystal (the layers in the 1D system considered here, as shown in Fig. \ref{fig:fig1}) have itself a nonlinear response. Therefore, for a sufficiently high amplitude, nonlinearity and periodicity act together at any point in the material, differently from other proposals where the periodic and nonlinear parts of the devices are disjunct. 
	 
In order to introduce the main ingredients of our proposal, we consider a bi-cromatic input with two close frequencies, such that the bandgaps of each component are both located close to one boundary of the crystal. In the nonlinear regime, the propagation of the two frequency components creates a low frequency wave by the self-demodulation effect. In addition, the dispersion introduced by the the chirped crystal, can be tuned to have bandgaps at the high frequency part of the wave field while the self-demodulated frequency lies in propagating regime, as schematically shown in Fig. \ref{fig:fig1}. When the bi-cromatic wave impinges on the crystal from left to right, both components can propagate along the crystal before reflected, while the self-demodulated wave is generated and transmitted (see Fig. \ref{fig:fig1}). However, propagation from the opposite side is different. Incident waves find firstly the bandgaps, and they are reflected without activating the self-demodulation effect as they do not propagate, and therefore there is no transmittance at all.  
This asymmetric propagation is a direct consequence of the nonlinear behaviour of the system, that breaks the time reversal symmetry and provides a non-reciprocal device.

The article is organized as follows. Section \ref{sec:structure} presents the chirped structure used for the investigation and describes the two configurations studied. The theoretical modeling based on a nonlinear wave equation and its numerical solution are described in Section \ref{sec:theoretical}. The linear properties of the system, as its dispersion relation, are described in Sec. \ref{sec:linear}. Section \ref{sec:results} shows numerical results demonstrating the asymmetric propagation, and gives an estimation of the efficiency of the process. Finally, some concluding remarks are discussed in Sec. \ref{sec:conclusions}.

\section{Chirped phononic crystal} 
\label{sec:structure}

The geometry of our proposed structure for an asymmetric propagation 
device is presented schematically in Fig. \ref{fig:fig1}.  
The one-dimensional (1D) system is composed of two homogeneous slabs of material 1, coupled to a chirped phononic crystal (CPC) placed between of them. The CPC is composed of $N$ layers of materials 2 and 3, placed alternatively at positions ranging from $x=L/2$ to $x=3L/2$, where $L$ represents the total length of the CPC. The spatial distribution of the chirped structure is given by the recurrence relation
\begin{eqnarray}
a_{i+1} &=& a_0 - e^{ \alpha x_{i}}, \nonumber\\
x_{i+1} &=& a_0e^{ \alpha x_{i}} + x_i,
\label{eq:chirped}
\end{eqnarray}
where $i=1,...,N$ is the number of layer, $a_i$ is the thickness of the \textit{i}-th layer, $a_0$ the first layer thickness, $\alpha$ the adimensional chirp parameter, and $x_i$ is the left boundary position of the $i$-th layer. The layer thickness  increases (decreases) in the positive direction of the $x$-axis, hence the chirped adimensional parameter is negative, $\alpha < 0$ ($\alpha > 0$). The $i$-th layer lattice parameter is given by the sum of the thicknesses of materials 2, 3: $a_i = a_{i,2} + a_{i,3}$, where $a_{i,2} = a_{i,3} = a_i/2$, i.e., the filling factor ($a_{i,2}/a_i$) is kept constant in each $i$-th layer. The chirped structure length is given by $L=\sum_{i=1}^Na_i$. We consider a CPC with the following values for the geometrical parameters: $a_0=0.12$ mm, $\alpha = -15.65$, and $N=500$, for a total length $L \approx 17,6$ cm. The structure presented in this work, is expected to be tested experimentally, therefore materials 1, 2, and 3, are selected on the basis of realistic materials as nickel-steel, aluminum and PMMA, respectively (nonlinear properties extracted from Ref. \onlinecite{Hamilton98Ch}). The physical properties of interest for this study are densities, $\rho_1 = 7950$ kg/m$^3$, $ \rho_2 = 2700$ kg/m\ $^3$, $\rho_3= 1180$ kg/m$^3$, speed of sound, $c_1 = 5790$ m/s, $c_2 = 6374$ m/s,$c_3 = 2680$ m/s, and the parameter of nonlinearity, $\beta_1 = 2.84$, $\beta_2 = 12.4$, $\beta_3 = 15.0$. We notice that we select high nonlinear materials for the CPC together with materials having low nonlinearity for the homogeneous slabs in order to reduce the nonlinear effect in the transmission lines out of the CPC and attribute all the nonlinear effects to the CPC and not to the propagation outside the CPC.
 
\section{Theoretical model}
\label{sec:theoretical}
For the sake of simplicity, in this work we consider a purely 1D system. The source velocity vector points in the $x$-direction, leading only to the generation of longitudinal modes. Under this consideration, the nonlinear acoustic modes in the solid material can be described by a second order nonlinear wave equation. In analogy with nonlinear compressional waves in fluids, and up to second order of accuracy, we model the full-wave propagation using the Westervelt equation\cite{naugolnykh1998nonlinear}, which in inhomogeneous media is written as
\begin{equation}\label{eq:westervelt}
 \frac{\partial ^2 p(x,t)}{\partial x^2} - \frac{1}{c_0^2(x)}\frac{\partial^2 p(x,t)}{\partial t^2} - \frac{1}{\rho _0(x)} \frac{\partial \rho _0(x)}{\partial x} \frac{\partial p(x,t)}{\partial x}  = - \frac{\beta(x)}{\rho _0(x) c_0^4(x)}\frac {\partial^2 p^2(x,t) }{\partial t^2},
\end{equation}
where $p(x,t)$ is the acoustic pressure, and $c_0(x)$, $\rho_0(x)$ and $\beta(x)$ are the space dependent sound speed, density and parameter of nonlinearity, respectively. Inside the multilayer structure these ambient properties are introduced as a space dependent function of alternating materials ($c_j$, $\rho_j$, $\beta_j$, $j=2,3$) following Eq.~(\ref{eq:chirped}). 

\section{Results}
\label{sec:results}
In this work we solve both the eigenvalue and the nonlinear propagation problems of the CPC previously defined. Although the CPC structure is not strictly periodic, if the chirp parameter $\alpha$ is sufficiently small, i.e., the layer thickness varies smoothly, the structure can be considered locally periodic, meaning that we can solve the eigenvalue problem at each position of the system considering that it behaves locally as an infinitely extended periodic crystal with local properties. Then, we can associate the spatial dispersion characteristics of the CPC, as demonstrated in Refs. \onlinecite{Romero13, Cebrecos14}.

In order to study the wave propagation inside the CPC, we will numerically solve the Eq.~(\ref{eq:westervelt}) by explicit finite differences using a computational grid of 200 elements per wavelength and a Courant-Friedrich-Levy number of 0.97. Two configurations are studied. First, in the forward configuration, ``+", the incoming wave is generated at the left boundary ($x=0$) and propagates in the positive direction of \textit{x}-axis (see Fig. \ref{fig:fig1}(a)). Second, for the reverse configuration, ``$-$", the incoming wave is generated at the right boundary ($x=2L$) and propagates in the negative direction of \textit{x}-axis (see Fig. \ref{fig:fig1}(b)). For both incidence directions the considered input signal is harmonic and composed of two slightly different frequencies, $f_1$ and $f_2$, having an amplitude $p_0 =p_{0,1}+p_{0,2}$, where $p_{0,1}=p_{0,2}=p_0/2$. We notice that for sufficiently small values of $p_0$, the propagation is linear. 
	
\subsection{System characterization in linear regime}
\label{sec:linear}

\begin{SCfigure}[1][htbp]
	\includegraphics[width=9 cm]{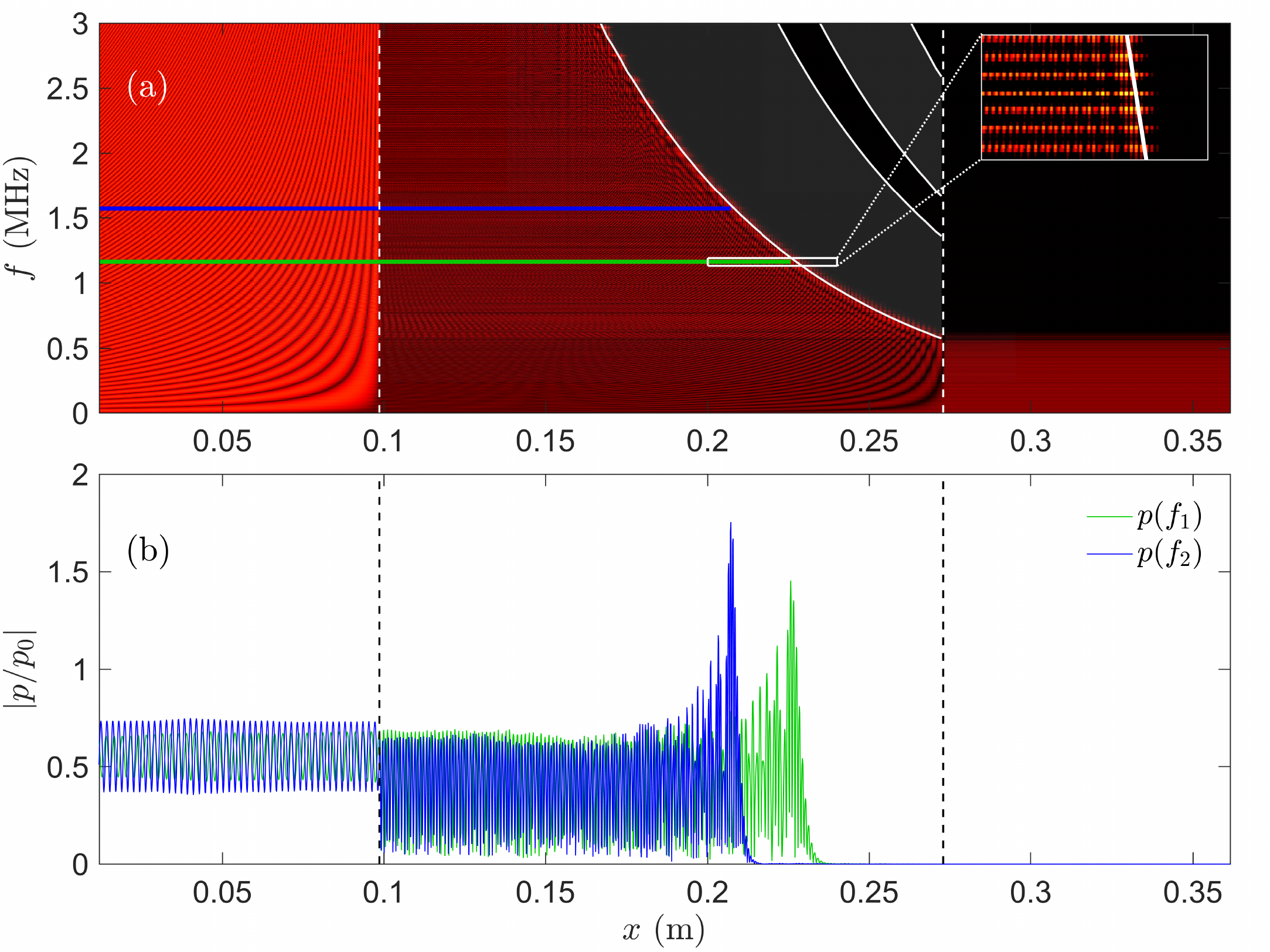}
	\caption{(a) Space-frequency representation of the acoustic field in linear regime. Green and blue lines highlight the selected pair of frequencies $f_1$, $f_2$. Solid white lines represent lower and upper edges of the spatially dependent bandgaps. Small inset zooms in a certain space-frequency region to highlight the existence of numerous Fabry-Perot resonances.  (b) Absolute value of the normalized pressure field in steady state for the frequencies of interest $f_1$, $f_2$. Dashed white (black) lines in top (bottom) inset indicate the interfaces between the CPC and the slabs of material 1.}
	\label{fig:fig2}
\end{SCfigure}

We begin describing the system behaviour in linear regime for the forward configuration. The CPC is designed in order to fulfill three main conditions: (1) The two components of the input signal have been selected in such a way that their frequencies fall inside  local bandgaps along their propagation through the CPC, (2) these local bandgaps are located nearly at the end of the structure, aiming the input wave propagates deep inside the structure to increase the interaction of both frequency components in the nonlinear regime, (3) the frequency of the self-demodulated wave is smaller than the cut-off frequency of the CPC. By cut-off frequency we mean the smallest frequency of the lower-edge of the spatially varying local bandgaps (see white lines in Fig. \ref{fig:fig2}(a)). The local dispersion relations of every layer along the CPC are calculated analytically using the well-known Rytov formula considering the $e^{\imath\omega t}$ harmonic dependence \cite{Kosevich06}
\begin{eqnarray}
\cos(q_ia_i)=\cos(k_2a_{i,2})\cos(k_3a_{i,3})- \frac{1}{2} \big( \frac{Z_{2}}{Z_{3}} + \frac{Z_{3}}{Z_{2}} \big) \sin(k_2a_{i,2})\sin(k_3a_{i,3}), 
\label{rytov}
\end{eqnarray} 
where $q_i$ is the Bloch wavenumber corresponding to a periodic media made by a unit cell as the $i$-th layer, $k_j = \omega / c_j$, where $c_j$ is the speed of sound in the layer of material $j=$ 2, 3, and $Z_{j}=\rho_j c_j$ is the acoustic impedance. Figure \ref{fig:fig2}(a) shows the space-frequency representation of the acoustic field up to 3 MHz. An impulse signal is applied in the left boundary ($x=0$), propagating through material 1 until its penetration into the CPC, where the enhancement of the pressure field takes place for the resonant frequencies at different depths (rainbow trapping). Solid white lines indicate the lower and upper edges of the spatially dependent local bandgaps, calculated using Eq. \ref{rytov}. Throughout this work, the selected frequencies of the input signal in nonlinear regime are $f_1=1.16$ MHz $f_2=1.57$ MHz (highlighted in blue and green color in Fig. \ref{fig:fig2}(a)). The pressure field in steady state for these frequencies is enhanced after its propagation along most part of the CPC, reaching the maximum value at different depths, as shown in Fig. \ref{fig:fig2}(b).  

\subsection{System characterization in nonlinear regime. Forward configuration.}
\label{subsec:DFG}

\begin{SCfigure}[1][htbp]
	\includegraphics[width=9 cm]{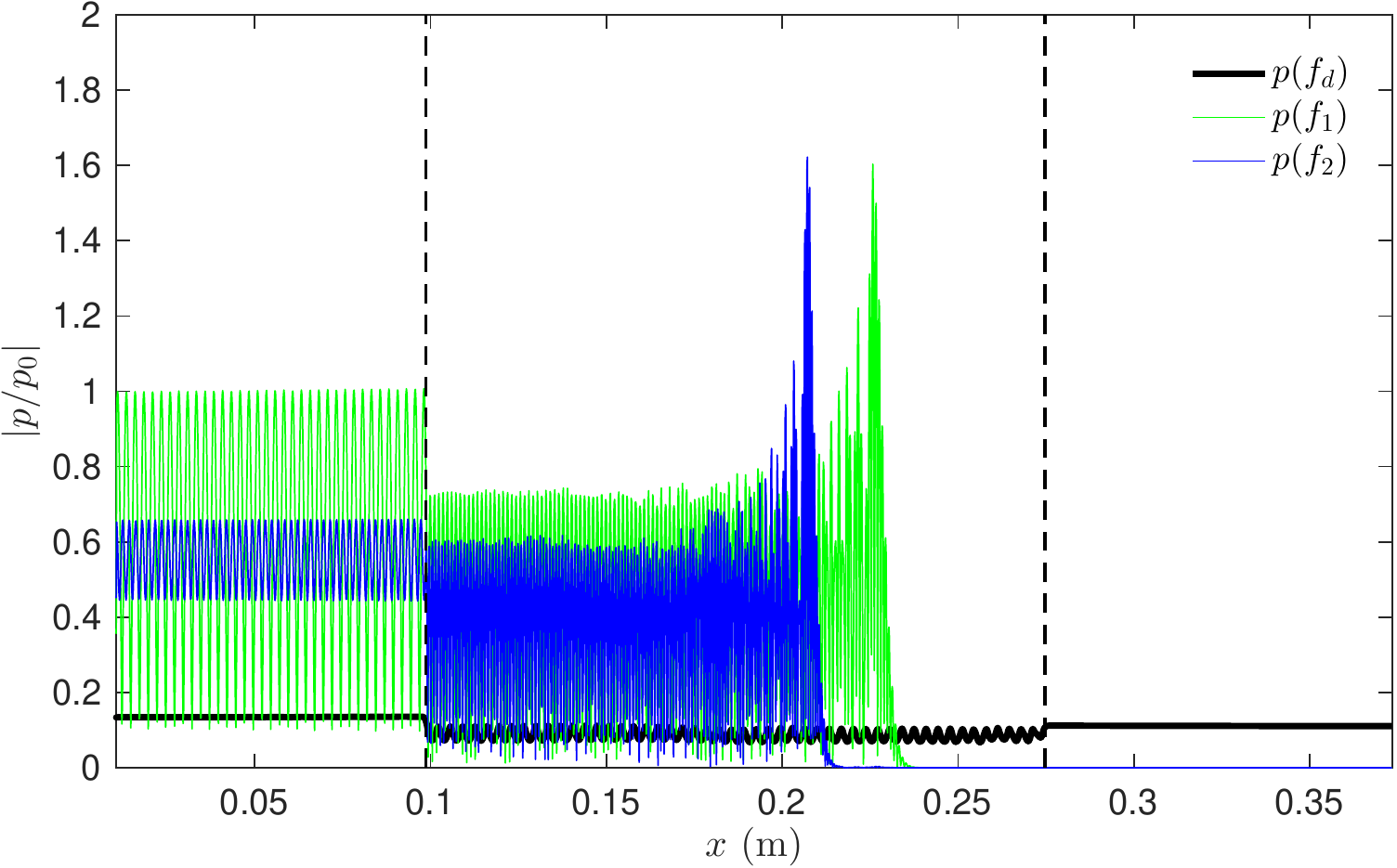}
	\caption{Harmonic distribution of the pressure field for the forward configuration in nonlinear regime, normalized to the amplitude of the incoming wave, $p_0= 8.5$ MPa. Dashed black lines highlight interfaces between the CPC and homogeneous slabs of material 1.}
	\label{fig:fig3}
\end{SCfigure}

The linear characterization shows how, for the selected input frequencies, the field is enhanced first and almost totally reflected later. We turn now our attention into the nonlinear self-demodulation process for the forward configuration. The pressure field distribution is shown in Fig. \ref{fig:fig3} for input frequencies $f_1$, $f_2$, and amplitude $p_0 = 8.5$ MPa. The self-demodulated wave has a frequency  $f_d= f_2-f_1 = 0.39$ MHz, which lies below the cut-off frequency of the CPC. The normalized amplitude of the difference-frequency wave at $x=2L$ is $|p/p_0| = 0.11$, while the amplitude of frequencies $f_1$, $f_2$, is strongly attenuated, having a normalized amplitude $|p/p_0| \approx 10^{-5}$, for each frequency component. Recalling the filtering feature of the designed CPC, not only the input frequency components $f_1$, $f_2$, are filtered out from the transmitted wave, but also their higher harmonics, as well as any linear combination between them above the cut-off frequency of the CPC. It is worthy to note that the enhancement of the pressure field in the CPC is the result of a resonant process inside the structure. As a consequence, it results necessary to let the propagating wave to reach the steady state. In this regard, it is noted that the pressure field distribution shown in Fig. \ref{fig:fig3} is calculated using a time window corresponding to 20 cycles of the demodulated signal, once the steady state is reached.

\subsection{Asymmetric propagation}
\label{subsec:asymmetric}

\begin{figure}[htbp]
	\includegraphics[width=14 cm]{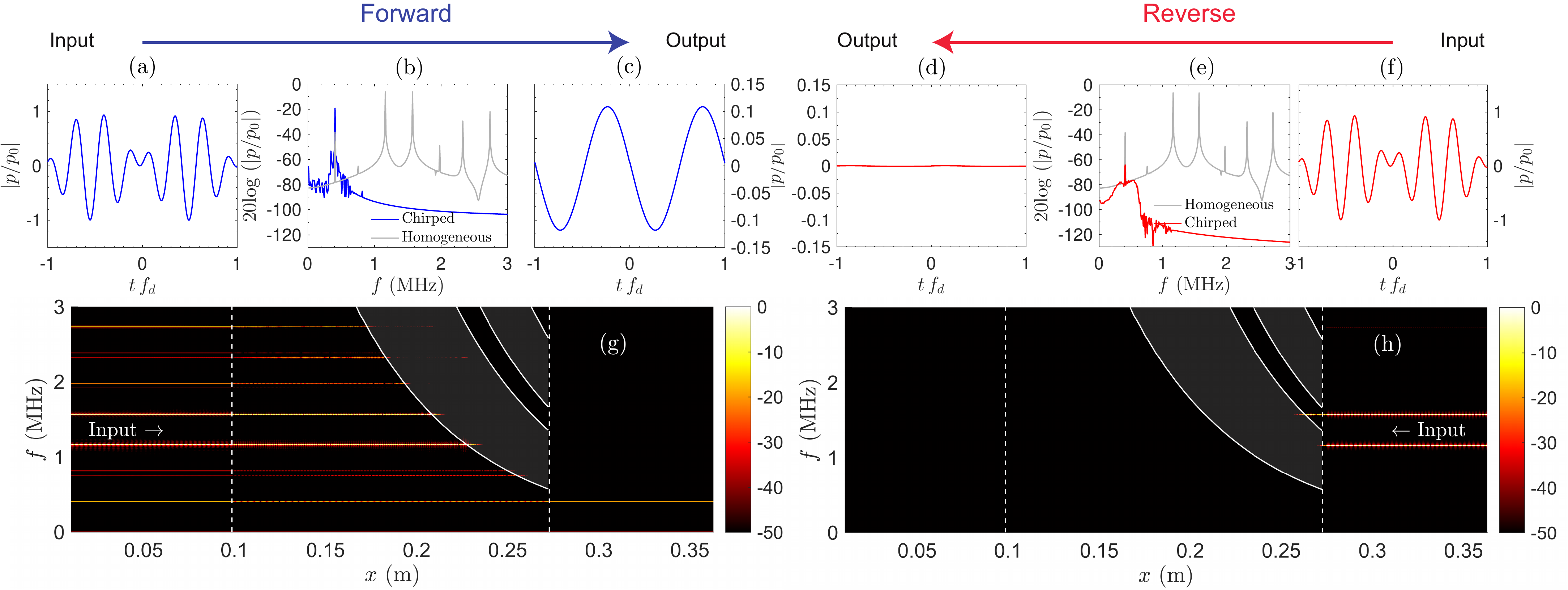}
	\caption{Behaviour of the asymmetric propagation device in ``+" (blue) and ``$-$" (red) directions. (a), (f) Input signal applied in $x=0$ ($x=2L$) m in ``+" (``$-$") direction. (b), (e) Spectrum of the output time signals in $x=2L$ ($x=0$) m in ``+" (``$-$") direction. Grey solid line represent the spectrum obtained for a system where the CPC is substituted by an homogeneous medium of material 2. (c), (d) Output time signals in ``+" and ``$-$"  directions, respectively. (g), (h) Space-frequency distribution of the normalized intensity in ``+" and ``$-$" directions, respectively. Vertical white dashed lines indicate the boundaries of the CPC.}
	\label{fig:fig4}
\end{figure}

The whole picture of the results obtained using the proposed asymmetric propagation device is shown in Fig. \ref{fig:fig4} for both forward (blue) and reverse (red) configurations, including all the relevant time signals (Figs. \ref{fig:fig4}(a, c) for ``+" direction, Figs. \ref{fig:fig4}(d, f) for ``$-$" direction), spectrum at the exit of the CPC (Figs. \ref{fig:fig4}(b, e)) and the harmonic distribution of the acoustic field in space (Figs. \ref{fig:fig4}(g,  h)). The most relevant feature of the device is its asymmetric and non-reciprocal character, i.e., the ability to generate and transmit the difference-frequency component of the signal on one direction while not in the other. However, the device performance goes far beyond this main goal. 

Firstly, the transmitted signal in the ``+" direction, originated from the self-demodulation process, is efficiently generated, having a normalized pressure amplitude $|p/p_0| = 0.11$ at $x=2L$. In order to show that the origin of this efficient generation is related to the enhancement of the field inside the CPC, a secondary simulation is performed: the CPC is substituted by an homogeneous slab composed of material 2 (aluminum). Comparing the spectrum shown in Fig. \ref{fig:fig4}(b), where the normalized intensity at $x=2L$ is shown in logarithmic scale, $I_+=20\log(|p/p_0|)$ , it is observed a nearly 20 dB reduction in the intensity of the demodulated wave in the homogeneous case (-19 dB using CPC versus -38 dB using the homogeneous medium). 

Secondly, every frequency component above the cut-off frequency of the CPC is completely filtered out from the output signal in both configurations (see Figs. \ref{fig:fig4}(b, e)), with intensities between -75 and -105 dB in ``+" direction and -85 and -125 dB in ``$-$" direction. We notice that the spatial characteristics of the CPC offer greater options when selecting the input frequencies $f_1$, $f_2$, compared to a homogeneous slab of the same length. Fabry-P\'erot resonances in the CPC are not equi-spaced in frequency, as velocity depends on it, $f_n = nc(f)/2L$. Hence, the demodulated signal frequency can be modified just by selecting another pair of resonances whose difference lie below the cut-off frequency.

To quantify the asymmetry of the device, we make use of the energy rectification ratio \cite{liang2010acoustic, boechler2011bifurcation, devaux2015asymmetric}, which is defined here as  $\sigma = I(+) / I(-)$ in terms of the forward and reverse intensities, which are proportional to the acoustic energy. For the specific case shown in this Section, the obtained energy rectification ratio is $\sigma = 3 \times 10^4$, which is of the same order of other solutions proposed in the literature \cite{liang2010acoustic,boechler2011bifurcation}.   

\subsection{Amplitude dependence}
\label{subsec:Idependence}

\begin{SCfigure}[1][htbp]
	\includegraphics[width=9 cm]{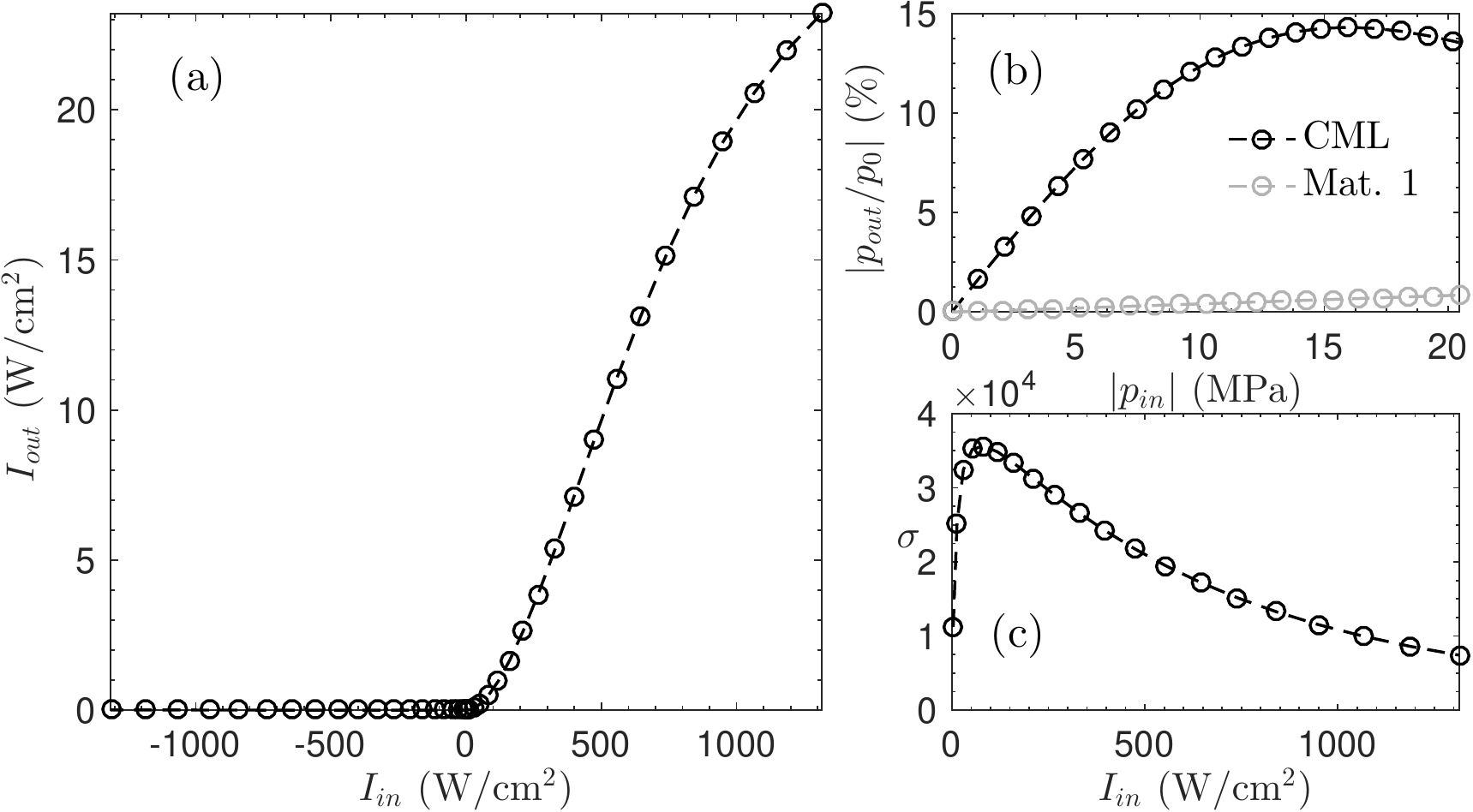}
	\caption{(a) Effect of the input intensity $I_{in}$ on the output intensity for both ``-" and ``+" directions. The input amplitude ranges from 0 to 20 MPa (0 to 1300 (W/cm$^{2}$)) for each direction of propagation. (b) Efficiency of the system in terms of pressure in ``+" direction, compared to an homogeneous medium made of material 2. (c) Rectification ratio in terms of the input intensity.}
	\label{fig:fig5}
\end{SCfigure}

The amplitude dependence of the non-reciprocal device is calculated in terms of the input and output intensities, $I_{in}$, $I_{out}$, being $I= p^2 / 2 \rho_1 c_1$. It is noted that due to the effective filtering performed by the CPC, the output signal, both in ``+" and ``$-$" directions, consists basically in the difference-frequency component, although in the latter case the amplitude of the wave is very small. Figure \ref{fig:fig5}(a) shows the asymmetric character of the proposed system. For negative values of the intensity the transmitted energy is almost zero, while for positive values the transmitted energy increases approximately in a linear manner for small and moderate values of the input intensity. 

In the previous Section we have seen that the enhancement of the acoustic field inside the chirped structure results in a more efficient self-demodulation process if compared to an homogeneous medium (see Figs. \ref{fig:fig4}(b, e)). Here, the efficiency of the self-demodulation process in ``+" direction in terms of the input pressure is also studied. Figure \ref{fig:fig5}(b) shows that for every value of the input pressure the efficiency of the proposed device is considerably higher than the one obtained placing an homogeneous slab of material 1 instead of the CPC. The amplitude dependence of the pressure shows different regimes. For small amplitude values of the input pressure, below 8 MPa, the output pressure increases linearly. As the amplitude increases the efficiency of the self-demodulation process is reduced as the energy of the input harmonics is transferred to higher harmonics. For the highest values, this energy transfer results in a slight reduction of the output pressure and as a consequence in a reduction of the efficiency. 

In order to show the ranges of efficiency of our system, the dependence of the energy rectification ratio values on the input intensity are also studied and shown in Fig. \ref{fig:fig5}(c). The obtained values of the rectification ratio are of the order of $10^4$ for all the inputs used in this work. This means that  the asymmetry and non-reciprocity character of the CMT structure is robust in terms of the intensities of the input signal, allowing a wide operational range. 

\section{Conclusions} 
\label{sec:conclusions}         

A novel design of an asymmetric and non-reciprocal structure for acoustic propagation based on the nonlinear self-demodulation effect has been theoretically presented. The structure consist of a chirped phononic crystal made of nonlinear acoustic media. The numerical study performed here reveals that the structure plays a double role of frequency conversion and filtering. The enhancement of the field inside the CPC plays a key role to increase the efficiency in ``+" direction compared to an homogeneous medium. The spatially varying dispersion, characteristic of chirped structures, allows to filter out the input frequency components and their higher harmonics, achieving strong attenuation values. This filter character of our structure has a direct consequence in the corresponding values of the rectification ratio, achieving a value of $\approx 10^4$ for values of the input intensity that ranges between $[200, 1300]$ $W/cm^2$. The interplay between local periodicity and nonlinearity is revealed as a useful tool for the design of devices with asymmetric transmission properties. 

\begin{acknowledgments}

This work was supported by Ministerio de Econom\'a y Competitividad (Spain) and European Union FEDER through project FIS2015-65998-C2-2-P, and by PROPASYM project funded by the R\'egion Pays-dela-
Loire. 

\end{acknowledgments}


%

\end{document}